\documentclass[journal, twoside]{IEEEtran}
\usepackage[utf8]{inputenc}
\usepackage[per-mode=symbol]{siunitx}
\DeclareSIUnit{\sample}{Sa}
\DeclareSIUnit{\baud}{Bd}
\DeclareSIUnit{\bit}{b}
\DeclareSIUnit{\byte}{B}

\usepackage{hyperref}
\hypersetup{hidelinks}

\usepackage[acronym,nomain]{glossaries}
\glsdisablehyper

\usepackage{comment}
\usepackage{lipsum}
\usepackage{graphicx}
\usepackage{grffile}
\usepackage{booktabs}
\usepackage{cite}
\usepackage{nicefrac}
\usepackage{balance}
\usepackage[caption=false,font=footnotesize]{subfig}
\usepackage{multirow}
\usepackage{float}
\usepackage{placeins} 
\usepackage{stfloats}
\usepackage{pdfpages}
\usepackage[capitalise]{cleveref}
\usepackage{amsmath}
\usepackage{amssymb}
\newcommand{\SetCapsType}{normalcaps}

\makeatletter
\providecommand{\SetCapsType}{smallcaps}

\long\def\@scTrue{smallcaps}
\long\def\@scFalse{normalcaps}
\newcommand{\acroSCaps}[1]{%
 \begingroup
  \ifx\SetCapsType\@scTrue 
    \textsc{#1}%
  \else
    \MakeUppercase{#1}%
  \fi
  \endgroup
}
\makeatother

\newcommand{\nAcronym}[4][]{%
	\newacronym[#1]{#2}{#3}{#4}
}

\makeatletter
\@ifpackageloaded{babel}{%
    \newcommand{\usuk}[2]{%
        \iflanguage{USenglish}{#1}{#2}%
    }%
}{%
    \newcommand{\usuk}[2]{%
        #1%
    }%
}%
\makeatother

\usepackage[shortcuts]{extdash}
\newcommand{\qam}[1]{
    \ifglsused{QAM}%
        {#1\=/\gls{QAM}}%
        {#1\=/ary \gls{QAM}%
    }%
}%

\newcommand{\pam}[1]{
    \ifglsused{PAM}%
        {#1\=/\gls{PAM}}%
        {#1\=/ary \gls{PAM}%
    }%
}%

\nAcronym{2A8PSK}{\acroSCaps{2a8psk}}{2-ary amplitude 8-ary phaseshift keying}

\nAcronym{3CCMCF}{\acroSCaps{3cc-mcf}}{3-core coupled-core multi-core fiber}

\nAcronym{4D}{\acroSCaps{4d}}{four-dimensional}
\nAcronym{4D64PRS}{\acroSCaps{4d-64prs}}{\usuk{four-dimensional 64-ary polarization-ring-switching}{four-dimensional 64-ary polarisation-ring-switching}}

\nAcronym{5B4D2A8PSK}{\acroSCaps{5b4d-2a8psk}}{5-bit four-dimensional two-amplitude 8-ary phase-shift keying}

\nAcronym{6B4D2A8PSK}{\acroSCaps{6b4d-2a8psk}}{6-bit four-dimensional two-amplitude 8-ary phase-shift keying}

\nAcronym{8D}{\acroSCaps{8d}}{eight-dimensional}\nAcronym{8D2048PRS}{\acroSCaps{8d-2048prs}}{eight-dimensional 2048-ary polarization-ring-switching}
\nAcronym{8D2048PRST1}{\acroSCaps{8d-2048prs-t1}}{eight-dimensional 2048-ary polarization-ring-switching type 1}
\nAcronym{8D2048PRST2}{\acroSCaps{8d-2048prs-t2}}{eight-dimensional 2048-ary polarization-ring-switching type 2}

\nAcronym{ABC}{\acroSCaps{abc}}{automatic bias controller}
\nAcronym{ADC}{\acroSCaps{adc}}{analog-to-digital converter}
\nAcronym{AIR}{\acroSCaps{air}}{achievable information rate}
\nAcronym{AOM}{\acroSCaps{aom}}{acoustic optical modulator}
\nAcronym{AR}{\acroSCaps{ar}}{achievable rate}
\nAcronym{ASE}{\acroSCaps{ase}}{amplified spontaneous emission}
\nAcronym{ASIC}{\acroSCaps{asic}}{application-specific integrated circuit}
\nAcronym{AWGN}{\acroSCaps{awgn}}{additive white Gaussian noise}

\nAcronym{BER}{\acroSCaps{ber}}{bit error rate}
\nAcronym{BICM}{\acroSCaps{bicm}}{bit-interleaved coded modulation}
\nAcronym{BMD}{\acroSCaps{bmd}}{bit-metric decoding}
\nAcronym{BPD}{\acroSCaps{bpd}}{balanced photo-diode}
\nAcronym{BPS}{\acroSCaps{bps}}{blind phase search}
\nAcronym{BRGC}{\acroSCaps{brgc}}{binary reflected Gray code}
\nAcronym{BTB}{\acroSCaps{btb}}{back-to-back}

\nAcronym{CCDM}{\acroSCaps{ccdm}}{constant composition distribution matching}
\nAcronym{CCF}{\acroSCaps{ccf}}{\usuk{coupled-core fiber}{coupled-core fibre}}
\nAcronym{CD}{\acroSCaps{cd}}{chromatic dispersion}
\nAcronym{CIR}{\acroSCaps{cir}}{channel impulse response}
\nAcronym{CMUX}{\acroSCaps{cmux}}{core multiplexer}
\nAcronym{ChUT}{\acroSCaps{chut}}{channel under test}
\nAcronym{CUT}{\acroSCaps{cut}}{channel under test}
\nAcronym{CPE}{\acroSCaps{cpe}}{carrier phase estimation}
\nAcronym{CAGR}{\acroSCaps{cagr}}{compound annual growth rate}

\nAcronym{DA}{\acroSCaps{da}}{driver amplifier}
\nAcronym{DAC}{\acroSCaps{dac}}{digital-to-analog converter}
\nAcronym{DCF}{\acroSCaps{dcf}}{\usuk{dispersion compensated fiber}{dispersion compensated fibre}}
\nAcronym{DCI}{\acroSCaps{dci}}{data center interconnect}
\nAcronym{DFB}{\acroSCaps{dfb}}{distributed feedback}
\nAcronym{DGD}{\acroSCaps{dgd}}{differential group delay}
\nAcronym{DM}{\acroSCaps{dm}}{distribution matcher}
\nAcronym{DMGD}{\acroSCaps{dmgd}}{differential mode group delay}
\nAcronym{DP}{\acroSCaps{dp}}{\usuk{dual-polarization}{dual-polarisation}}
\nAcronym{DPC}{\acroSCaps{dpc}}{digital pre-compensation}
\nAcronym{DPE}{\acroSCaps{dpe}}{digital pre-emphasis}
\nAcronym{DPIQ}{\acroSCaps{dp-iqm}}{\usuk{dual-polarization IQ-modulator}{dual-polarisation IQ-modulator}}
\nAcronym{DRE}{\acroSCaps{dre}}{digital resolution enhancer}
\nAcronym{DSO}{\acroSCaps{dso}}{digital sampling oscilloscope}
\nAcronym{DUT}{\acroSCaps{dut}}{device-under-test}
\nAcronym{DWDM}{\acroSCaps{dwdm}}{dense wavelength-division multiplexing}

\nAcronym{ED}{\acroSCaps{ed}}{Eucledian distance}
\nAcronym{ENOB}{\acroSCaps{enob}}{effective number of bits}
\nAcronym{ESS}{\acroSCaps{ess}}{enumerative sphere shaping}

\nAcronym{FDE}{\acroSCaps{fde}}{\usuk{frequency domain equalizer}{frequency domain equaliser}}
\nAcronym{FEC}{\acroSCaps{fec}}{forward error correction}
\nAcronym{FFT}{\acroSCaps{fft}}{fast Fourier transform}
\nAcronym{FWM}{\acroSCaps{fwm}}{four-wave mixing}

\nAcronym{GFF}{\acroSCaps{gff}}{gain flattening filter}
\nAcronym{GIMMF}{\acroSCaps{gi-mmf}}{\usuk{graded-index multi-mode fiber}{graded-index multi-mode fibre}}
\nAcronym{GMI}{\acroSCaps{gmi}}{\usuk{generalized mutual information}{generalised mutual information}}
\nAcronym{GS}{\acroSCaps{gs}}{geometric shaping}
\nAcronym{GV}{\acroSCaps{gv}}{group-velocity}
\nAcronym{GVD}{\acroSCaps{gvd}}{group-velocity dispersion}

\nAcronym{HDFEC}{\acroSCaps{hd-fec}}{hard decision forward error correction}

\nAcronym[plural=IL, firstplural=insertion losses (\textsc{il})]{IL}{\acroSCaps{il}}{insertion loss}
\nAcronym{IFFT}{\acroSCaps{ifft}}{inverse fast Fourier transform}
\nAcronym{ISI}{\acroSCaps{isi}}{intersymbol interference}

\nAcronym{KK}{\acroSCaps{kk}}{Kramers-Kronig}

\nAcronym{LCOS}{\acroSCaps{LCoS}}{liquid crystal on silicon}
\nAcronym{LDPC}{\acroSCaps{ldpc}}{low-density parity-check}
\nAcronym{LEAF}{\acroSCaps{leaf}}{\usuk{large effective area fiber}{large effective area fibre}}
\nAcronym{LLR}{\acroSCaps{llr}}{log-likelihood ratio}
\nAcronym{LO}{\acroSCaps{lo}}{local oscillator}
\nAcronym{LSPS}{\acroSCaps{lsps}}{\usuk{loop-synchronized polarization scrambler}{loop-synchronised polarisation scrambler}}
\nAcronym{LUT}{\acroSCaps{lut}}{lookup table}

\nAcronym{MB}{\acroSCaps{mb}}{Maxwell-Bolzmann}
\nAcronym{MDM}{\acroSCaps{mdm}}{mode division multiplexing}
\nAcronym{MF}{\acroSCaps{mf}}{matched filter}
\nAcronym{MI}{\acroSCaps{mi}}{mutual information}
\nAcronym{MMA}{\acroSCaps{mma}}{multi-modulus algorithm}
\nAcronym{MPLC}{\acroSCaps{mplc}}{multi-plane light converter}
\nAcronym{MSE}{\acroSCaps{mse}}{mean squared error}
\nAcronym{MUX}{\acroSCaps{mux}}{multiplexer}

\nAcronym{NF}{\acroSCaps{nf}}{noise figure}
\nAcronym{NGMI}{\acroSCaps{ngmi}}{\usuk{normalized generalized mutual information}{normalised generalised mutual information}}
\nAcronym{NIR}{\acroSCaps{nir}}{near infra-red}

\nAcronym{OBTB}{\acroSCaps{obtb}}{optical back-to-back}
\nAcronym{ODL}{\acroSCaps{odl}}{optical delay line}
\nAcronym{OFDR}{\acroSCaps{ofdr}}{optical frequency-domain reflectometer}
\nAcronym{OFDM}{\acroSCaps{ofdm}}{orthogonal frequency division multiplexing}
\nAcronym{OMFT}{\acroSCaps{omft}}{optical-multi-format transmitter}
\nAcronym{OSA}{\acroSCaps{osa}}{\usuk{optical spectrum analyzer}{optical spectrum analyser}}
\nAcronym{OTDR}{\acroSCaps{otdr}}{optical time-domain reflectometer}
\nAcronym{OTF}{\acroSCaps{otf}}{optical tunable filter}
\nAcronym{OVNA}{\acroSCaps{ovna}}{\usuk{optical vector network analyzer}{optical vector network analyser}}

\nAcronym{PAM}{\acroSCaps{pam}}{pulse-amplitude modulation}
\nAcronym{PAS}{\acroSCaps{pas}}{probabilistic amplitude shaping}
\nAcronym{PAPR}{\acroSCaps{papr}}{peak-to-avarage power ratio}
\nAcronym{PBS}{\acroSCaps{pbs}}{polarization beam splitter}
\nAcronym{PCVD}{\acroSCaps{pcvd}}{plasma chemical vapor depostion}
\nAcronym{PDL}{\acroSCaps{pdl}}{polarization-dependent loss}
\nAcronym{PDM}{\acroSCaps{pdm}}{\usuk{polarization-division multiplexing}{polarisation-division multiplexing}}
\nAcronym{PMBPSK}{\acroSCaps{pm-bpsk}}{polarization-multiplexed binary phase-shift-keying}
\nAcronym{PMQPSK}{\acroSCaps{pm-qpsk}}{polarization-multiplexed quaternary phase-shift-keying}
\nAcronym{PM8QAM}{\acroSCaps{pm-8qam}}{polarization-multiplexed 8-ary quadrature amplitude modulation}
\nAcronym{PMD}{\acroSCaps{pmd}}{polarization mode dispersion}
\nAcronym{PNOB}{\acroSCaps{pnob}}{physical number of bits}
\nAcronym{PRBS}{\acroSCaps{prbs}}{pseudorandom bit sequence}
\nAcronym{PS}{\acroSCaps{ps}}{probabilistic shaping}

\nAcronym{RF}{\acroSCaps{rf}}{radio frequency}
\nAcronym{SamPerSym}{\acroSCaps{sps}}{samples per symbol}
\nAcronym{SDFEC}{\acroSCaps{sd-fec}}{soft decision forward error correction}
\nAcronym{SE}{\acroSCaps{se}}{spectral efficiency}
\nAcronym{SER}{\acroSCaps{ser}}{symbol error rate}
\nAcronym{SA}{\acroSCaps{sa}}{simulated annealing}
\nAcronym{SLM}{\acroSCaps{slm}}{spatial light modulator}
\nAcronym{SSFM}{\acroSCaps{ssfm}}{split-step Fourier method}
\nAcronym{SMF}{\acroSCaps{smf}}{\usuk{single-mode fiber}{single-mode fibre}}
\nAcronym[firstplural=\usuk{states of polarization (\acroSCaps{sop})}{states of polarisation (\acroSCaps{sop})}]{SOP}{\acroSCaps{sop}}{\usuk{state of polarization}{state of polarisation}}
\nAcronym{SPM}{\acroSCaps{spm}}{self-phase modulation}
\nAcronym{SPS}{\acroSCaps{sps}}{\usuk{synchronized polarization scrambler}{synchronised polarisation scrambler}}
\nAcronym{SSMF}{\acroSCaps{ssmf}}{\usuk{standard single-mode fiber}{standard single-mode fibre}}
\nAcronym{SVD}{\acroSCaps{svd}}{singular value decomposition}
\nAcronym{SWI}{\acroSCaps{swi}}{swept wavelength interferometry}

\nAcronym{TDE}{\acroSCaps{tde}}{\usuk{time domain equalizer}{time domain equaliser}}
\nAcronym{TDMSDM}{\acroSCaps{tdm-sdm}}{time-domain multiplexed space-division multiplexing}
\nAcronym{TE}{\acroSCaps{TE}}{transverse electric}
\nAcronym{TM}{\acroSCaps{TM}}{transverse magnetic}
\nAcronym{TH4D}{\acroSCaps{th-4d}}{time domain hybrid four-dimensional}
\nAcronym{TH4D2A8PSK}{\acroSCaps{th-4d-2a8psk}}{time domain hybrid four-dimensional two-amplitude eight-phase shift keying}

\nAcronym{VOA}{\acroSCaps{voa}}{variable optical attenuator}

\nAcronym{WSS}{\acroSCaps{wss}}{wavelength selective switch}

\nAcronym{XPM}{\acroSCaps{xpm}}{cross-phase modulation}
\nAcronym{XT}{\acroSCaps{xt}}{cross-talk}

\nAcronym{3DWG}{\acroSCaps{3dwg}}{3D-waveguide}

\nAcronym{DD}{\acroSCaps{dd}}{decision-directed}
\nAcronym{DMD}{\acroSCaps{dmd}}{differential-mode delay}
\nAcronym{DSP}{\acroSCaps{dsp}}{digital signal processing}

\nAcronym{ECL}{\acroSCaps{ecl}}{external cavity laser}
\nAcronym{EDFA}{\acroSCaps{edfa}}{\usuk{erbium-doped fiber amplifier}{erbium-doped fibre amplifier}}

\nAcronym{GD}{\acroSCaps{gd}}{group delay}
\nAcronym{GIFMF}{\acroSCaps{gi-fmf}}{\usuk{graded-index few-mode fiber}{graded-index few-mode fibre}}

\nAcronym{LP}{\acroSCaps{lp}}{\usuk{linearly polarized}{linearly polarised}}
\nAcronym{LMS}{\acroSCaps{lms}}{least mean square}
\nAcronym{MMSE}{\acroSCaps{mmse}}{minimum mean square error}
\nAcronym{MCF}{\acroSCaps{mcf}}{\usuk{multi-core fiber}{multi-core fibre}}
\nAcronym{MDG}{\acroSCaps{mdg}}{mode-dependent gain}
\nAcronym{MDL}{\acroSCaps{mdl}}{mode-dependent loss}
\nAcronym{MMF}{\acroSCaps{mmf}}{\usuk{multi-mode fiber}{multi-mode fibre}}
\nAcronym{MZM}{\acroSCaps{mzm}}{Mach-Zehnder modulator}
\nAcronym{MIMO}{\acroSCaps{mimo}}{multiple-input multiple-output}

\nAcronym{FMF}{\acroSCaps{fmf}}{\usuk{few-mode fiber}{few-mode fibre}}
\nAcronym{FM-MCF}{\acroSCaps{fm-mcf}}{\usuk{few-mode multi-core fiber}{few-mode multi-core fiber}}

\nAcronym{OSNR}{\acroSCaps{osnr}}{optical signal-to-noise ratio}

\nAcronym{PL}{\acroSCaps{pl}}{photonic lantern}

\nAcronym{RRC}{\acroSCaps{rrc}}{root-raised-cosine}

\newacronym{SDM}{SDM}{space division multiplexing}
\nAcronym{SNR}{\acroSCaps{snr}}{signal-to-noise ratio}

\nAcronym{ULH}{\acroSCaps{ulh}}{ultra-long-haul}

\nAcronym{QAM}{\acroSCaps{qam}}{quadrature amplitude modulation}
\nAcronym{QPSK}{\acroSCaps{qpsk}}{quaternary phase-shift-keying}

\nAcronym{WDM}{\acroSCaps{wdm}}{wavelength-division multiplexing}

\nAcronym{SINR}{\acroSCaps{sinr}}{signal-to-interference-plus-noise ratio}

\nAcronym{LS}{\acroSCaps{ls}}{least-squares}

\nAcronym{ANN}{\acroSCaps{ann}}{artificial neural network}

\nAcronym{ML}{\acroSCaps{ml}}{machine learning}

\nAcronym{CombIR}{\acroSCaps{cir}}{combined impulse response}

\usepackage[]{xcolor}

\newcommand{\LP}[1]{LP\textsubscript{#1}}
\newcommand{\eig}[1]{%
    \ifthenelse { \equal {#1} {} }%
    {\def\temp{}}%
    {\def\temp{,\textrm{#1}}}%
    $\lambda^{2}_{i\temp}$%
}

\usepackage{silence}
\makeatletter
\let\blx@rerun@biber\relax
\makeatother

\begin{document}
\bstctlcite{IEEEexampleLimitAuthors:BSTcontrol}
%


    \title{MDG and SNR Estimation in SDM Transmission Based on Artificial Neural Networks}

%
%
%

    \author{Ruby~S.~B.~Ospina,~\IEEEmembership{Student Member,~OSA,}
        Menno~van~den~Hout,~\IEEEmembership{Student Member,~IEEE,}\\
        Sjoerd~van~der~Heide,~\IEEEmembership{Student Member,~IEEE,}
        John~van~Weerdenburg,~\IEEEmembership{Student Member,~IEEE,}\\
        Roland~Ryf,~\IEEEmembership{Fellow,~IEEE,}
        Nicolas~K.~Fontaine,~\IEEEmembership{Senior Member,~IEEE,}
        Haoshuo~Chen,
        Rodrigo~Amezcua-Correa,~\IEEEmembership{Member,~IEEE},
        ~Chigo~Okonkwo,~\IEEEmembership{Senior Member,~IEEE,}\\
        and Darli~A.~A.~Mello,~\IEEEmembership{Member,~IEEE}%
        \thanks{Manuscript received XXX xx, XXXX; revised XXXXX xx, XXXX; accepted XXXX XX, XXXX. Date of publication XXXX XX, XXXX. This work was partially supported by FAPESP under grants 2018/25414-6, 2018/14026-5, 2017/25537-8, 2015/24341-7, 2015/ 24517-8 and by the TU/e-KPN Smart Two project. This article was presented in part at the Optical Fiber Communications Conference, San Diego, CA, USA, June. 2021.}
        \thanks{R. S. B. Ospina and D. A. A. Mello are with the School of Electrical and Computer Engineering, State University of Campinas, Campinas 13083-970, Brazil (e-mail: ruby@decom.fee.unicamp.br, darli@unicamp.br).}%
        \thanks{M. van den Hout, Sjoerd van der Heide, and C. Okonkwo are with the High Capacity Optical Transmission Laboratory, Electro-Optical Communications Group, Eindhoven University of Technology, PO Box 513, 5600 MB, Eindhoven, The Netherlands. (e-mail: \{m.v.d.hout; s.p.v.d.heide; c.m.okonwko\}@tue.nl).}%
        \thanks{J. van Weerdenburg was formerly with Eindhoven University of Technology, and is now with Infinera, San Jose, CA 95119 USA (e-mail: jweerdenburg@infinera.com). }%
        \thanks{R. Ryf, N. K. Fontaine, and H. Chen are with NokiaBell Labs, Holmdel, NJ 07733 USA (e-mail: \{roland.ryf; nicolas.fontaine; haoshuo.chen\}@nokia-bell-labs.com).}%
        \thanks{R. Amezcua-Correa is with the CREOL, The College of Optics and Photonics, University of Central Florida, Orlando, 32816, USA (e-mail: r.amezcua@creol.ucf.edu)}
        \thanks{Color versions of one or more figures in this paper are available online at http://ieeexplore.ieee.org.}
        \thanks{Digital Object Identifier xxxxxxxxxxxxxxxxxxxxxxxx}
    }

    \IEEEpubid{}

    \IEEEpubid{\begin{minipage}{\textwidth}
                   \ \\[12pt]
                   \centering
                   0000--0000/00\$00.00~\copyright~2021 IEEE Personal use is permitted, but republication/redistribution requires IEEE permission.\\See http://www.ieee.org/publications standards/publications/rights/index.html for more information.
    \end{minipage}}



    \maketitle

    \begin{abstract}
        The increase in capacity provided by coupled \gls{SDM} systems is fundamentally limited by \gls{MDG} and \gls{ASE} noise. Therefore, monitoring  MDG and optical \gls{SNR} is essential for accurate performance evaluation and troubleshooting. Recent works show that the conventional MDG estimation method based on the transfer matrix of \gls{MIMO} equalizers optimizing the \gls{MMSE} underestimates the actual value at low \glspl{SNR}. Besides, estimating the optical SNR itself is not a trivial task in SDM systems, as MDG strongly influences the electrical SNR after the equalizer.  In a recent work we propose an MDG and SNR estimation method using \glspl{ANN}. The proposed \gls{ANN}-based method processes features extracted at the receiver after \gls{DSP}. In this paper, we discuss the ANN-based method in detail, and validate it in an experimental 73-km 3-mode transmission link with controlled MDG and SNR. After validation, we apply the method in a case study consisting of an  experimental long-haul 6-mode link. The results show that the \gls{ANN} estimates both MDG and SNR with high accuracy, outperforming conventional methods.
        
    \end{abstract}

    \begin{IEEEkeywords}
        Mode-dependent loss, mode-dependent gain, space division multiplexing, optical fiber communications.
    \end{IEEEkeywords}

    \glsreset{SDM}
    \glsreset{ASE}
    \glsreset{MDL}
    \glsreset{MDG}
    \glsreset{SNR}
    \glsreset{MIMO}
    \glsreset{MMSE}
    \glsreset{DSP}
    \glsreset{ANN}

%
    \IEEEpeerreviewmaketitle

    \section{Introduction}
    
   \Gls{SDM} with coupled channels is a promising solution to scale the fiber capacity in future optical system generations.
    Coupled \gls{SDM} has been effectively demonstrated in laboratory experiments over coupled-core \glspl{MCF} \cite{ryf2019coupled}, \glspl{MMF} \cite{fontaine201530}, \cite{van20201ecoc}, \glspl{FMF} \cite{van2018138}, \cite{rademacher2018long}, and \glspl{FM-MCF} \cite{soma201710}.\IEEEpubidadjcol \, Among the impairments that affect coupled \gls{SDM} transmission, the interaction of additive noise and \gls{MDG}\footnote{The results of this paper apply to the combined effects of  \gls{MDG} and \gls{MDL}. However, for the sake of simplicity, we refer simply to  \gls{MDG}.} fundamentally limits the system capacity. The random power variations of guided modes induced by  \gls{MDG} turn the channel capacity into a random variable, reducing the average capacity and generating outages \cite{ho2011mode,winzer2011mimo,8918466}. Therefore, assessing the accumulated \gls{MDG} and the optical \gls{SNR} at the receiver is essential for performance evaluation and troubleshooting.

      In the recent literature, \gls{MDG} has been estimated by \gls{DSP} using the transfer function of \gls{MIMO} \cite{van2018138,rademacher2018long,van2017138, rademacher202010} equalizers. 
    However, we show in  \cite{ospina2020dsp} that, as adaptive \gls{MIMO} equalizers typically use the \gls{MMSE} criterion \cite{faruk2017digital}, the DSP-based estimation accuracy is affected by noise. We show that the accumulated \gls{MDG} is underestimated for high levels of MDG and low optical SNRs\footnote{We avoid using the OSNR acronym because we evaluate the optical SNR at the signal bandwidth, instead of the usual \SI{12.5}{\giga\hertz} bandwidth.}. To circumvent this limitation, we propose in \cite{ospina2020mode} a correction factor to partially compensate for \gls{MDG} estimation errors. The validity of the correction factor is verified experimentally in \cite{ospina2020mode,van2020experimental}. One drawback of the correction factor is that it requires a known optical \gls{SNR} that may not be readily available.

    Estimating the optical SNR in coupled SDM receivers is also not trivial. In single-mode transmission, polarization-dependent gain (PDG) is not a limiting effect, and the optical SNR can be estimated from the electrical SNR by a simple direct equation \cite{essiambre2010capacity}.
    In coupled SDM transmission, however, the electrical SNR may be strongly affected by \gls{MDG}. In this case, estimating the optical SNR directly from the electrical SNR would underestimate the actual value.

    Currently, \gls{ML} techniques are being extensively investigated for optical performance monitoring in both single-mode \cite{saif2020machine} and mode-multiplexed systems \cite{saif2020optical}. In \cite{ospina2021ann}, we propose an \gls{ANN}-based solution to estimate both \gls{MDG} and optical \gls{SNR} in coupled SDM transmission. The results are validated in a back-to-back \SI{32.5}{m}  3-mode \gls{FMF} link. This paper extends the results in \cite{ospina2021ann}, discussing the method in detail, and  validating it in an experimental short-haul \SI{73}{km} 3-mode \gls{FMF} link with controlled \gls{MDG} and optical \gls{SNR}. In addition, we apply the \gls{ANN} estimator in a case study of an experimental long-haul 6-mode  transmission setup with unknown \gls{MDG} and \gls{SNR}.
    
    The remainder of this paper is structured as follows. Section II reviews the conventional methods used to estimate MDG and SNR in coupled SDM transmission. Section III presents the \gls{ANN}-based solution. Section IV presents  validation results in an experimental short-reach transmission setup. Section V applies the method in a long-haul case study. Lastly, Section VI concludes the paper.

    \begin{figure*}[t]
    \centering
    \includegraphics[width=1\linewidth]{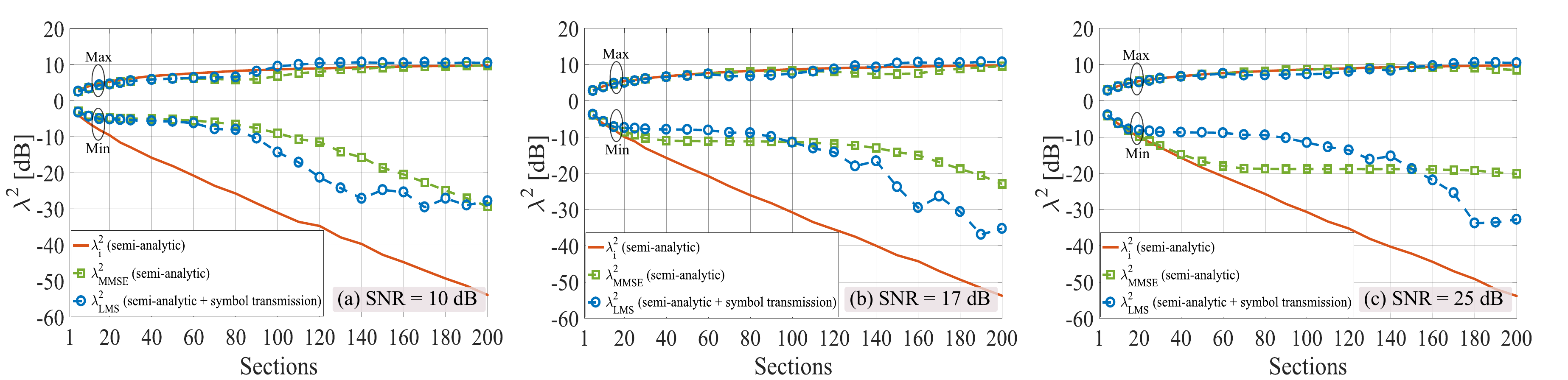}%
    \caption{{Evolution of the maximum and minimum actual eigenvalues, $\lambda_{i}^{2}$, and DSP-estimated eigenvalues, $\lambda^{2}_{{\mathrm{MMSE}}}$, and $\lambda^{2}_{{\mathrm{LMS}}}$, with the number of sections. (a) At low SNR = 10 dB. (b) At medium SNR = 17 dB. (c) At high SNR = 25 dB.}}
    \label{fig:error_vs_distance_conv_corrfactor_ann}
    \end{figure*}

    \section{Conventional methods for MDG and SNR estimation}
    \label{sec:conventionalestimation}

    \subsection{MDG estimation}
    The \gls{MDG} of a link with transfer matrix $\mathbf{H}$ can be computed from the eigenvalues \eig{} of  $\mathbf{H}\mathbf{H}^{H}$, where  $(.)^{H}$ is the Hermitian transpose operator \cite{ho2011mode, winzer2011mimo}. The accumulated MDG can be quantified by two metrics. The first one is the peak-to-peak MDG given by the subtraction of the largest and the lowest eigenvalues in dB ($10\textrm{log}_{10}{(\lambda_{i}^{\textrm{max}})^{2}}-10\textrm{log}_{10}{(\lambda_{i}^{\textrm{min}})^{2}}$) \cite{winzer2011mimo}. The second one is the standard deviation of the eigenvalues in logarithmic scale ($\sigma_{\mathrm{mdg}} = \mathrm{std(log(}\lambda_{i}^{2}))$). An interesting advantage of the standard deviation metric is that, in long-haul links with strong mode coupling, it allows to estimate the impact of \gls{MDG} on capacity using analytic formulas  \cite{ho2011mode}, \cite{8918466}. Therefore, in this paper, we use the standard deviation metric.

    In \gls{DSP}-based \gls{MDG} estimation, $\mathbf{H}$ is unknown. Alternatively, the inverse of the equalizer transfer function,  $\mathbf{W}^{-1}_{\mathrm{EQ}}$, is conventionally used as an estimate of $\mathbf{H}$ \cite{van2018138,rademacher2018long,rademacher202010}. MIMO receivers are usually implemented by MMSE equalizers, whose transfer function can be expressed as \cite{kim2008performance,mckay2009achievable}
    \begin{equation}
        \mathbf{W}_{\mathrm{MMSE}} = \left( \frac{\mathrm{\mathbf{I}}}{\mathrm{SNR}} + \mathbf{H}^{H}\mathbf{H} \right)^{-1}\mathbf{H}^{H},
        \label{Eq: wmmse}
    \end{equation}
    
    \noindent where $\mathrm{SNR}$  is calculated in optical domain using the signal bandwidth as reference noise bandwidth. The standard deviation 
    $\sigma_{\mathrm{mdg}}$ is then computed from the eigenvalues $\lambda^{2}_{i_{\mathrm{MMSE}}}$ of  $\mathbf{W}^{-1}_{\mathrm{MMSE}}(\mathbf{W}^{-1}_{\mathrm{MMSE}})^H$.

    From the eigendecomposition of  $\mathbf{W}^{-1}_{\mathrm{MMSE}}(\mathbf{W}^{-1}_{\mathrm{MMSE}})^H$, the relationship between the actual eigenvalues, \eig{}, and the eigenvalues obtained by \gls{DSP}, ${\lambda}^{2}_{i,\textrm{MMSE}}$, is given by \cite{ospina2020dsp}
    \begin{equation}
         \lambda^{2}_{i,\textrm{MMSE}} =  \left[\frac{\left(\lambda^{2}_{i}\right)^{-1}}{\mathrm{SNR}^{2}} + \frac{2}{\mathrm{SNR}} + \lambda^{2}_{i} \right].
        \label{Eq:Lambdarelation}
    \end{equation}
         The standard deviation metric is then estimated as
    
    \begin{equation}
        \widehat{\sigma}_{\mathrm{mdg}} = \mathrm{std(log(}\lambda_{i,\textrm{MMSE}}^{2})). \label{Eq:std_hat}
    \end{equation}

    \noindent \cref{Eq:Lambdarelation,Eq:std_hat} indicate that the accuracy of the conventional method that estimates $\sigma_{\mathrm{mdg}}$ based on the DSP-estimated eigenvalues, \eig{MMSE}, is clearly affected by the optical $\mathrm{SNR}$ \cite{ospina2020dsp}. In coupled SDM transmission, the MMSE equalizer is usually implemented  by means of semi-supervised or supervised adaptive schemes, such as the well-known \gls{LMS} algorithm. Although the LMS algorithm reaches the MMSE for mild channel conditions, it can suffer from implementation issues in extreme channel conditions, such as in pathological levels of MDG.
    

    \cref{fig:error_vs_distance_conv_corrfactor_ann} shows the maximum and minimum eigenvalues for coupled SDM transmission, for three different SNR values. For each transmission distance, matrices $\mathbf{H}$ are generated using the semi-analytical multisection model presented in \cite{ho2011mode}. The model simulates the coupled transmission of $2\mathrm{M} = 12$ spatial and polarization modes over 50-km fiber spans. The per-amplifier MDG standard deviation, $\sigma_{g}$, is set to 1 dB. Eigenvalues \eig{} are calculated directly from $\mathbf{H}$. Eigenvalues \eig{MMSE} are computed by inverting  $\mathbf{W}_{\mathrm{MMSE}}$ calculated in \cref{Eq: wmmse}. Eigenvalues \eig{LMS} are obtained by Monte-Carlo simulation of a complete coupled SDM transmission system. The transmitter generates 12 independent sequences of 460,000 16-QAM symbols at \SI{30}{\giga\baud}. The complex signals are shaped by \gls{RRC} filters and converted to the optical domain by a Mach-Zehnder modulator (MZM) model. The simulated channel consists of 1,000 frequency bins spread over 240 GHz (note that the simulation bandwidth is 30 GHz times 8 samples per symbol, yielding 240 GHz). The resolution of the channel in frequency domain is adjusted by replicating channel matrices between simulated frequency bins.
    The modal dispersion per span is 21.9 ps, corresponding to 50-km of a fiber with group delay standard deviation of 3.1 ps/$\sqrt{\mathrm{km}}$ \cite{hayashi2017record}. Additive white Gaussian noise is added with equal power to all received channel streams to set the desired receiver SNR. At the receiver, the signals are converted to the electric domain by a coherent receiver front-end model. The electrical signals are then filtered, digitized, and processed by the \gls{DSP} chain, including a fully-supervised \gls{LMS} algorithm.
    The detailed description of the simulation setup is presented in \cite{ospina2020mode}.

    In Figs. \ref{fig:error_vs_distance_conv_corrfactor_ann}(a)-(c), three regimes of \eig{MMSE} can be identified. 
    In the first regime, the absolute values of both maximum and minimum \eig{MMSE} simultaneously increase, tracking the actual eigenvalues \eig{}.
    In the second regime, both maximum and minimum \eig{MMSE} remain approximately constant, leading the minimum \eig{MMSE} to deviate considerably from \eig{}. In the third regime, the absolute value of the minimum \eig{MMSE} increases again. The LMS MIMO equalizer results in a maximum \eig{LMS} that tracks the maximum \eig{} and \eig{MMSE} with high accuracy over the entire link. On the other hand, the minimum \eig{LMS} diverges from the minimum \eig{MMSE}  for high values of accumulated MDG.

    The results in  \cref{fig:error_vs_distance_conv_corrfactor_ann} indicate that eigenvalues derived directly from the equalizer coefficients, such as  \eig{LMS}, track the actual eigenvalues \eig{} only over low-MDG links. For long distances and high MDG, conventional estimation methods largely underestimate the link MDG. The correction factor proposed in \cite{ospina2020dsp} can partially compensate for this mismatch in scenarios of moderate MDG and low SNR, where \eig{MMSE} tracks \eig{LMS}, however, its correction capability is limited in pathological scenarios of extremely high MDG. Although these pathological scenarios may seem unlike in a first glance, it is possible to reach these levels in weakly coupled transmission, for which the MDG increases linearly with the link length.  
      \begin{figure*}[t]
	\centering
		\includegraphics[width=18.2cm]{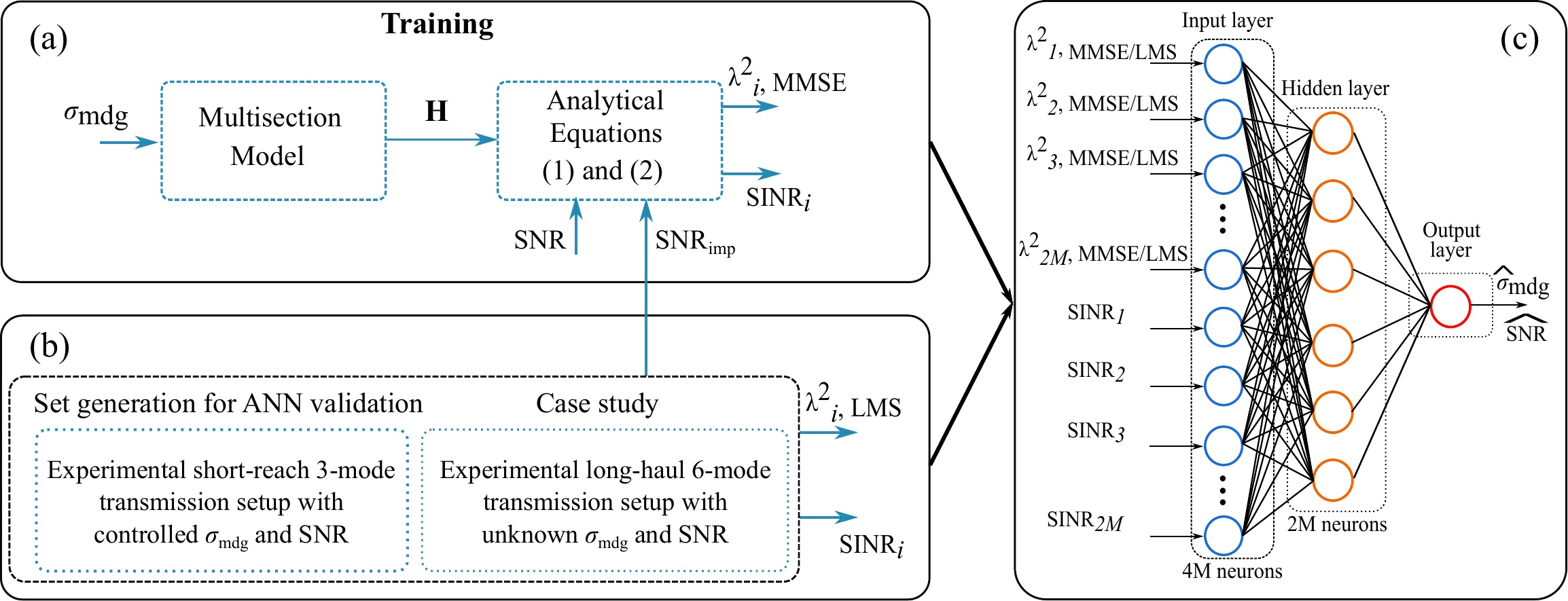}%
	\caption{{\gls{ANN}-based MDG and SNR estimator. (a) The training set is generated by numerical multisection simulation and analytic formulas. (b) The validation set is generated by a short-reach 3-mode transmission setup and the case study data is generated by a long-haul 6-mode transmission setup. (c)~Proposed ANN. The algorithm applies two separate networks for $\mathrm{\sigma_{mdg}}$ and SNR estimation.}}
	\label{fig:generaldiagram}
\end{figure*}

    \subsection{Optical SNR estimation}
    
    Estimating the optical \gls{SNR} is also not trivial in coupled SDM systems affected by MDG.  In systems with coherent detection, the optical \gls{SNR} can be estimated from the so-called electrical \gls{SNR}. In systems with MMSE equalization, the electrical \gls{SNR} in stream $i$ is actually a signal-to-noise-plus interference ratio ($\mathrm{SINR}_i$) \cite{mckay2009achievable}
    
    \begin{equation}
     \mathrm{SINR}_{i} = \frac{1}{\left[ \left( \mathbf{I} + \mathrm{SNR} \; \mathbf{H}^{H}\mathbf{H}  \right)^{-1} \right]_{i,\, i}} - 1,
    \label{Eq:SNRSINRrelation}
    \end{equation}
    where $[\;]_{i, \, i}$ indicates the i-th element in the main diagonal. The optical SNR is then estimated as

    \begin{equation}
        \mathrm{\widehat{SNR}} = \frac{1}{2M}\sum_{i=1}^{2M} \mathrm{SINR}_{i}. \label{snr:hat}
    \end{equation}

    In single-mode transmission with low PDG, $\mathbf{H}$ is approximately unitary, such that, in \cref{Eq:SNRSINRrelation} and \cref{snr:hat},  $\mathrm{\widehat{SNR}} \approx \mathrm{SNR}$. Therefore, $\mathrm{\widehat{SNR}}$ is usually obtained from $\mathrm{SINR}_{i}$, which is calculated using the \gls{LS} method \cite{wautelet2007comparison,das2012snr}. In MDG-impaired SDM systems, however, $\mathbf{H}$ is non-unitary, turning $\mathrm{\widehat{SNR}}$ dependent on $\mathbf{H}$. In this case, estimating the SNR  from the $\mathrm{SINR_i}$ would underestimate the actual value.
    
    \subsection{Implementation penalty}
    
    Another issue that may be taken into account in \cref{Eq: wmmse,Eq:Lambdarelation,Eq:std_hat,Eq:SNRSINRrelation,snr:hat} is the fact that implementation imperfections also affect the interplay of noise and MDG. These imperfections can be modeled as a contribution added to the optical noise. In this case, the  $\mathrm{SNR}$ can be redefined as $\mathrm{SNR'}$, expressed as
      \begin{equation}
        \mathrm{SNR'} = \left(\frac{1}{\mathrm{SNR}}+ \frac{1}{\mathrm{SNR_{imp}}}\right)^{-1},
        \label{Eq:newSNR}
    \end{equation}
    where $\mathrm{SNR_{imp}}$ is an implementation penalty computed as the average $\mathrm{SINR}_{i}$ estimated from the equalized data streams in back-to-back, i.e., without any MDG and optical noise.

 To improve the accuracy of the conventional SNR estimation technique that employs the \gls{LS} method, the implementation penalty contribution can be removed. In this case, the estimated SNR is redefined as
      \begin{equation}
         \mathrm{\widehat{SNR}} = \left(\frac{1}{\frac{1}{2M}\sum_{i=1}^{2M}\mathrm{SINR}_{i}}- \frac{1}{\mathrm{SNR_{imp}}}\right)^{-1}.
         \label{Eq:newsinr}
     \end{equation}
    
\begin{figure*}[t]
\centering
    	\includegraphics[width=1\linewidth]{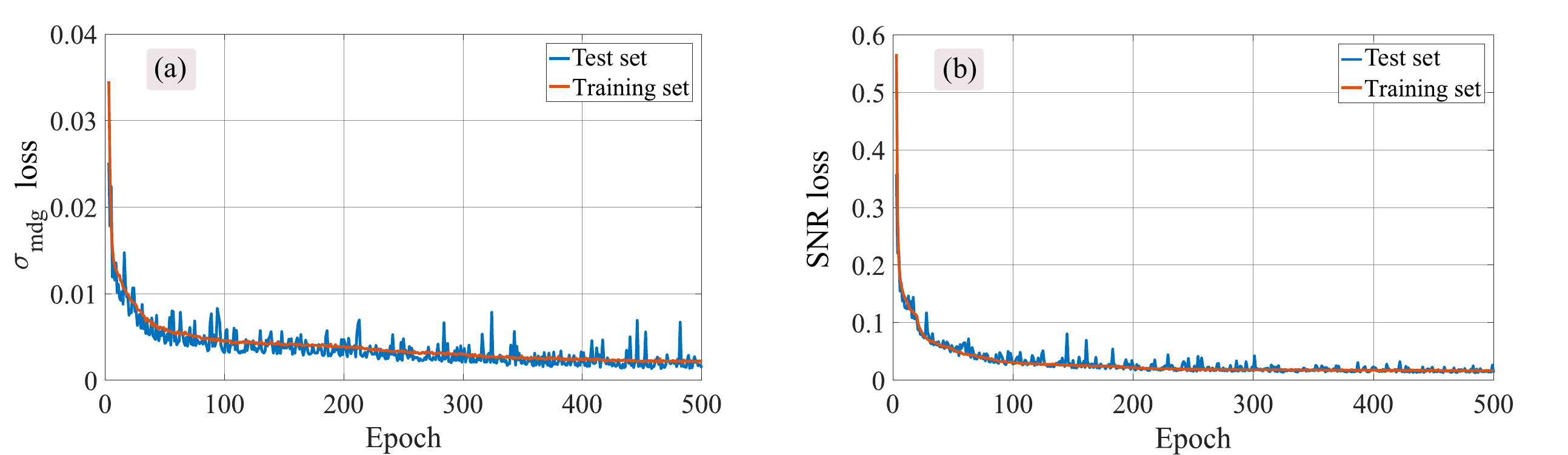}%
\caption{{ANN loss,  calculated as the \gls{MSE}, as a function of the number of epochs. (a) $\sigma_{\mathrm{mdg}}$ estimation. (b) SNR estimation. The curves indicate no overfitting and a good generalization ability.}}
\label{fig:ann_convergence}
\end{figure*}

   \section{ANN-based Method for MDG and SNR estimation}
   
To circumvent the limitations of conventional methods, we propose in \cite{ospina2021ann} an \gls{ANN}-based method to estimate $\mathrm{\sigma_{mdg}}$ and $\mathrm{SNR}$ from features extracted after DSP. The block diagram of the proposed solution is depicted in \cref{fig:generaldiagram}.

The training dataset is generated according to \cref{fig:generaldiagram}(a). Using the multisection model presented in \cite{ho2011mode}, $2\mathrm{M} \times 2\mathrm{M}$ matrices $\mathbf{H}$ are generated to simulate an $\mathrm{M}$-mode transmission  with polarization multiplexing over a link with $\mathrm{K} = $ 50 spans of 50 km each, yielding a total length of 2,500 km.

The overall \gls{MDG} of the link is controlled by the per-amplifier MDG standard deviation, $\sigma_{g}$. The standard deviation of the overall MDG is given by \cite{ho2011mode}

\begin{equation}
 \sigma_{\mathrm{mdg}} = \xi \sqrt{ 1+ \frac{\xi^{2}}{ 12 \, \left(1 - (\mathrm{2M})^{-2} \right) }},
\label{Eq:theosigmamdg}
\end{equation}

\noindent where the accumulated MDG standard deviation, $\xi$, increases with the number of spans, $\mathrm{K}$, as $\mathrm{\xi = \sigma_{\mathrm{g}}\; \sqrt{K}}$ \cite{ho2011mode}.  The MDG of the simulated 2,500 km-\gls{FMF} link is adjusted to result in  $\SI{0.2}{\decibel}<\mathrm{\sigma_{mdg}}<\SI{6.2}{\decibel}$.

For each $\mathbf{H}$, the $\mathrm{SNR}$ is swept from \SIrange{10}{25}{\decibel} to generate $2\mathrm{M}$ $\lambda^{2}_{i_{\mathrm{MMSE}}}$ values and $2\mathrm{M}$ $\mathrm{SINR}_{i}$ values using \cref{Eq: wmmse,Eq:SNRSINRrelation}. Input $\mathrm{SINR_{imp}}$, is a parameter measured experimentally in the transmission setup where the technique is being applied. The labelled set of $\lambda^{2}_{i_{\mathrm{MMSE}}}$ and $\mathrm{SINR}_{i}$ is fed into the \gls{ANN} shown in \cref{fig:generaldiagram}(c) as input training features. The \gls{ANN} receives $2\mathrm{M}$ $\lambda^{2}_{i_{\mathrm{MMSE}}}$ values and $2\mathrm{M}$ $\mathrm{SINR}_{i}$ values, and provides an estimate of $\sigma_{\mathrm{mdg}}$ or $\mathrm{SNR}$. A hidden layer with $2\mathrm{M}$ neurons, and an output layer with 1 neuron, learn the relation between the input features and the output. The \gls{ANN} is trained using the Adam optimizer \cite{kingma2014adam} using batches of 5 samples. 

Figs. \ref{fig:ann_convergence}(a) and \ref{fig:ann_convergence}(b) show the ANN convergence curves for MDG and SNR estimation, respectively, considering $2\mathrm{M} = 6$. The ANN loss, calculated as the \gls{MSE}, is depicted as a function of the number of epochs. Both training and test sample sets are evaluated.
The results indicate a substantial reduction in the MSE after 100 epochs, and still a small improvement up to 500 epochs for $\sigma_{\mathrm{mdg}}$ estimation. Therefore, we use 500 epochs for training. The loss of the test set tracks the loss of the training set for the entire figure. We expect, therefore, no overfitting and a good ANN ability to generalize over unseen samples.
After training, the \gls{ANN}-based method is validated using data captured from a short-reach experimental setup with controlled parameters, and tested in a case study of a long-haul experimental link.

\section{Experimental short-reach transmission}

 \subsection{Experimental short-reach 3-mode validation setup}
  \label{sec:setup}
\begin{figure*}[t!]  
        \vspace{-3mm}
        \centering
        \includegraphics[width=1\textwidth]{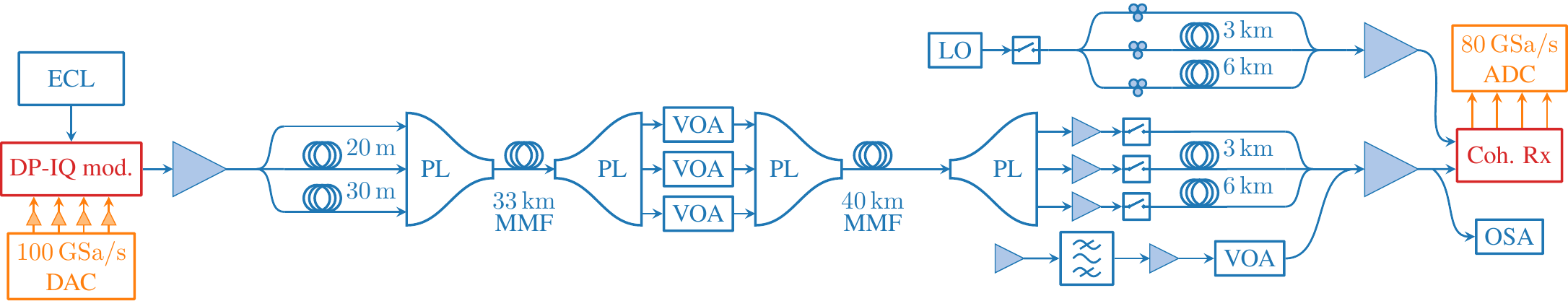}%
        \caption{Experimental setup for short-reach 3-mode transmission with polarization multiplexing \cite{van2020experimental}. The transmitter generates \qam{16} symbols at \SI{25}{\giga\baud}, which are subsequently split and delayed to create the input tributaries for the \gls{PL}. The multi-mode signal is transmitted over \SI{73}{\km} of \gls{MMF} \cite{sillard201650}. \glspl{VOA} are used to control the $\mathrm{\sigma_{mdg}}$ of the link. At the receiver, a \gls{TDMSDM} scheme is employed, and a noise loading stage is used to vary the \gls{OSNR}. After DSP, the LMS eigenvalues are computed from $\mathbf{W}^{-1}_{\mathrm{LMS}}(\mathbf{W}^{-1}_{\mathrm{LMS}})^H$. The $\mathrm{SINR}_{i}$ is computed from each of the 6 equalized data streams.}%
        \label{fig:expsetup}
    \end{figure*}

    The ANN-based estimator is validated using the short-reach 3-mode transmission setup presented in \cite{van2020experimental} and  depicted in \cref{fig:expsetup}. 
     The transmitter generates polarization-multiplexed \qam{16} symbols at a transmission rate of \SI{25}{\giga\baud}. A \gls{RRC} filter with 0.01 roll-off factor is used for pulse shaping. The pulse-shaped signal is converted to the analog domain by a 100 GSa/s \gls{DAC} followed by the optical modulator. After optical modulation, the signal is amplified by an \gls{EDFA}, split and delayed to generate three decorrelated data streams to be launched through the \LP{01}, \LP{11a} and \LP{11b} spatial modes. Considering polarization modes, the setup supports the transmission of 6 orthogonal modes. The three polarization-multiplexed data streams are then multiplexed in space by a mode-selective \acrfull{PL} \cite{velazquez2018scaling}.

     The output of the \gls{PL} is connected to a \SI{50}{\micro\meter} core diameter graded-index \gls{MMF} of \SI{73}{\km} \cite{sillard201650}. The deployed multi-mode fiber supports up to 36 spatial modes, so that transmission can be eventually scaled to more spatial modes. To control the overall MDG, two photonic lanterns and three \glspl{VOA} are placed after the first \SI{33}{\km} fiber segment. The three VOAs allow to sweep the MDG of the link by modifying the power in the three spatial modes. At the receive side, a fourth PL is used as mode de-multiplexer.

     The receiver employs a time-domain-multiplexed (TDM)-SDM receiver \cite{van2014ultra} to reduce the required amount of the coherent receivers. The ASE noise is varied at the coherent receiver input by a noise loading stage composed of two EDFAs, a wavelength selective switch (WSS) and a VOA. The $\mathrm{SNR}$ is computed as $\mathrm{ SNR = OSNR \, (T_{s} \times 12.5 \,GHz)}$, where $\mathrm{T_{s} = 40 \, ps}$ is the symbol time, and the OSNR is the traditional optical signal-to-noise ratio measured by an \gls{OSA} at the \SI{12.5}{\giga\hertz} bandwidth \cite{essiambre2010capacity}. The noisy signal is amplified and converted from the optical to the electrical domain by the receiver front-end. The TDM electric signals are fed into 80 GSa/s analog-to-digital converters (ADC) to be digitized.

     In the DSP block, the TDM streams are parallelized and down-sampled to two samples per symbol. To compensate for modal dispersion and linear coupling, 6$\times$6 MIMO equalization is carried out using a widely linear complex-valued adaptive equalizer, updated by a fully supervised \gls{LMS} algorithm \cite{da2016widely}. 
      After DSP, the eigenvalues \eig{LMS} are computed at each frequency of $\mathbf{W}_{\mathrm{LMS}}$ and averaged across the signal band. The $\mathrm{SINR}_{i}$ is computed from each of the 6 equalized data streams using a single-coefficient \gls{LS} estimator \cite{wautelet2007comparison}. The implementation penalty is computed in back-to-back as $\mathrm{SNR_{imp}}=\SI{18.8}{\decibel}$.

     The \gls{ANN} in \cref{fig:generaldiagram}(c) is fed with 9,610 analytical labelled samples generated using  \cref{Eq: wmmse,Eq:SNRSINRrelation} as indicated in \cref{fig:generaldiagram}(a). In a first stage, 8,649 samples are used for model training and the remainder 970 samples for model testing. After training, model validation is performed by 520 experimental samples generated by the short-reach 3-mode transmission setup. Using the \glspl{VOA} located in middle of the span,  $\mathrm{\sigma_{mdg}}$ is varied from \SIrange{4.5}{6.5}{\decibel}. At the receiver, the noise loading stage sweeps the $\mathrm{SNR}$ from \SIrange{11}{22}{\decibel}.
 
      \begin{figure*}[t]
\centering
    	\includegraphics[width=1\linewidth]{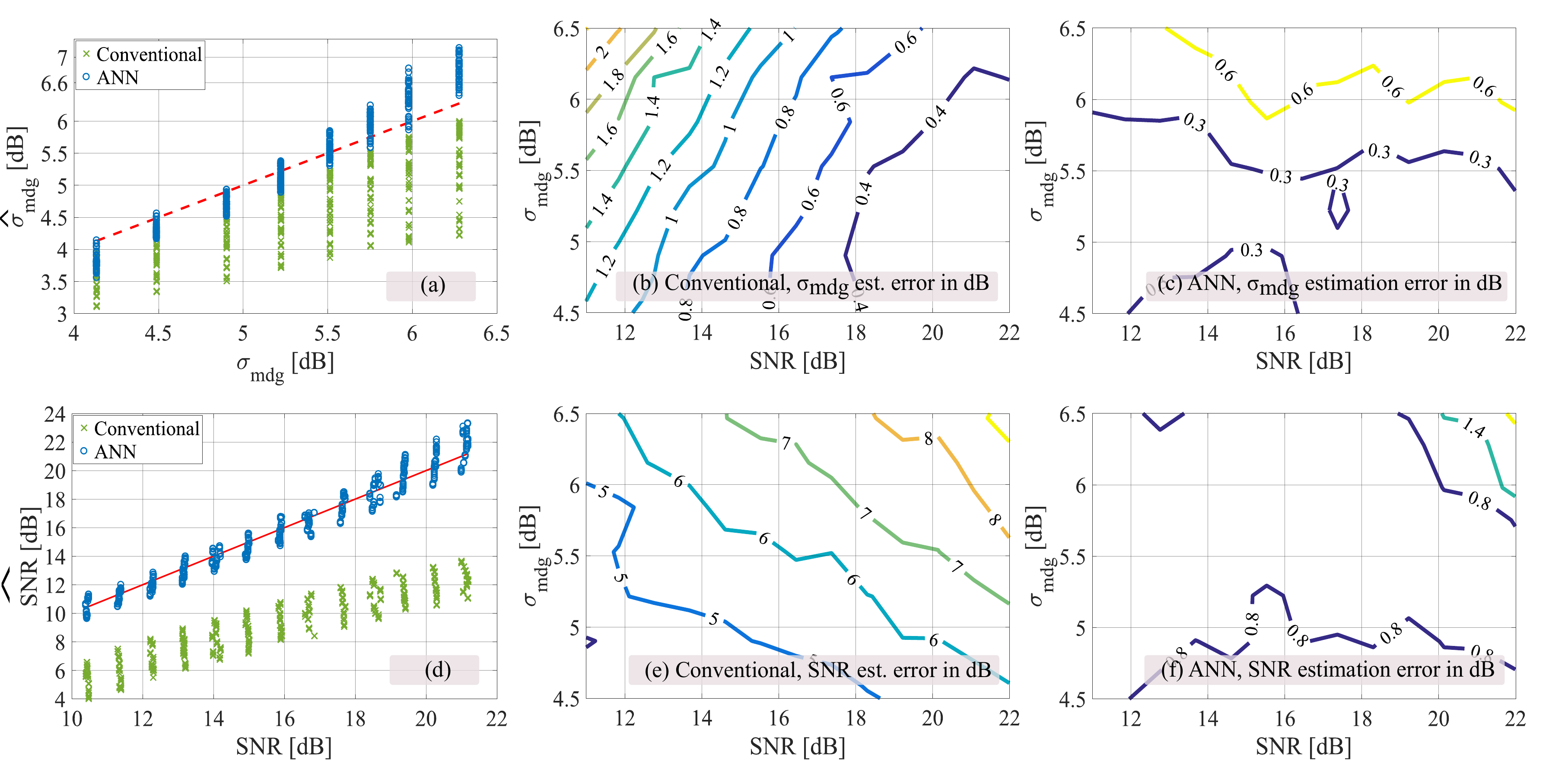}%
\caption{Experimental short-reach 3-mode validation results. (a) Estimated $\sigma_{\mathrm{mdg}}$ as a function of the actual $\sigma_{\mathrm{mdg}}$. (b) $\sigma_{\mathrm{mdg}}$ estimation error in dB generated by the conventional method as a function of the actual $\sigma_{\mathrm{mdg}}$ and $\mathrm{SNR}$. (c) $\sigma_{\mathrm{mdg}}$ estimation error in dB generated by the \gls{ANN} as a function of the actual $\sigma_{\mathrm{mdg}}$ and $\mathrm{SNR}$.
(d) Estimated $\mathrm{SNR}$ as a function of the actual $\mathrm{SNR}$. (e) $\mathrm{SNR}$ estimation error in dB generated by the conventional method as a function of the actual $\sigma_{\mathrm{mdg}}$ and $\mathrm{SNR}$. (f) $\mathrm{SNR}$ estimation error in dB generated by the \gls{ANN} as a function of the actual $\sigma_{\mathrm{mdg}}$ and $\mathrm{SNR}$.}
\label{fig:nn_pred_actual_73kminspan}
\end{figure*}

\subsection{Experimental short-reach 3-mode validation results}
\label{subsection:results_shortreach} 

 The validation results of the conventional and \gls{ANN}-based estimators in a short-reach transmission setup are depicted in \cref{fig:nn_pred_actual_73kminspan}.
 Figs. \ref{fig:nn_pred_actual_73kminspan}(a) and \ref{fig:nn_pred_actual_73kminspan}(d) compare the actual and estimated $\sigma_{\mathrm{mdg}}$ and $\mathrm{SNR}$ parameters generated by the conventional and  \gls{ANN}-based methods. The estimated values track the actual values within a small deviation over the entire range of parameters evaluated, resulting in a \gls{MSE} of 0.11 for $\sigma_{\mathrm{mdg}}$ and 0.53 for $\mathrm{SNR}$, both computed in dB.

Figs. \ref{fig:nn_pred_actual_73kminspan}(b) and \ref{fig:nn_pred_actual_73kminspan}(e) show the estimation error provided by the conventional method in dB, computed as the difference between the actual value and the estimated value.
The conventional method provides a $\sigma_{\mathrm{mdg}}$ estimation error up to \SI{2}{\decibel} at high MDG  and low $\mathrm{SNR}$. In the case of $\mathrm{SNR}$, the estimation error achieves up to \SI{8}{\decibel} at high levels of MDG  and high $\mathrm{SNR}$.
 Figs. \ref{fig:nn_pred_actual_73kminspan}(c) and \ref{fig:nn_pred_actual_73kminspan}(f) show the estimation error in decibels for $\sigma_{\mathrm{mdg}}$ and $\mathrm{SNR}$, respectively, for the \gls{ANN} solution.
The \gls{ANN} estimator provides a highest residual $\sigma_{\mathrm{mdg}}$ estimation error of \SI{0.6}{\decibel} in the region of high MDG, exhibiting a low dependence on the evaluated SNR. On most of the grid, the $\sigma_{\mathrm{mdg}}$ estimation error is lower than \SI{0.3}{\decibel}. For the $\mathrm{SNR}$, an estimation error higher than \SI{1.4}{\decibel} is observed at high values of $\sigma_{\mathrm{mdg}}$ and $\mathrm{SNR}$. Over most of the evaluated range, the SNR estimation error is lower than \SI{0.8}{\decibel}.

\section{Experimental long-haul transmission}

\subsection{Experimental long-haul 6-mode case study setup}
 
\begin{figure*}[t!]
        \centering
        \includegraphics[width=1\linewidth]{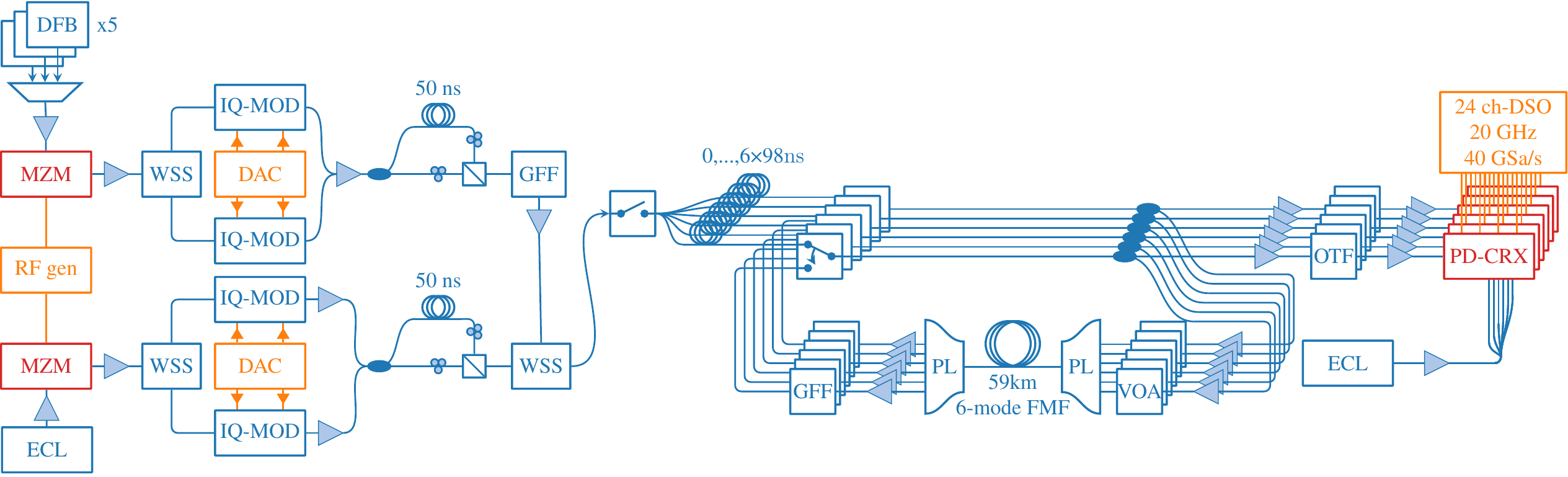}
        \caption{Experimental long-haul 6-mode transmission setup with recirculating loop of \SI{59}{\km}. 12 spatial and polarization modes are supported, each one carry 15 \gls{WDM} channels centered around 1550 nm. Triangles represent \glspl{EDFA}. 
        See \cite{van2018138} for more details.
        After DSP, the LMS eigenvalues are computed from $\mathbf{W}^{-1}_{\mathrm{LMS}}(\mathbf{W}^{-1}_{\mathrm{LMS}})^H$. The $\mathrm{SINR}_{i}$ is computed from each one of the 12 equalized data.} 
        \label{fig:setupbelllabs}
    \end{figure*}

We also apply the \gls{ANN}-based estimator to the long-haul 6-mode transmission with polarization multiplexing setup presented in \cite{van2018138} and depicted in  \cref{fig:setupbelllabs}.
The transmission setup includes 15 \gls{WDM} channels transmitted over 4 \gls{LP} spatial modes (\LP{01}, \LP{11}, \LP{21}, and \LP{02}). Including polarization and degenerate modes (\LP{11a}, \LP{11b}, \LP{21a}, and \LP{21b}) the setup supports 12 propagation modes.
 The 15-channel comb is generated using five distributed feedback lasers (DFB) and one phase-modulated Mach-Zehnder modulator (MZM). Odd and even channels are separately modulated with \SI{120}{\giga\bit\per\second} \qam{16} using IQ-modulators. Polarization-multiplexing is generated by splitting, delaying and combining the transmitted signals. The \gls{CUT} is generated separately using a similar scheme. Six conventional single-mode recirculating loops are combined with \glspl{PL} and a \SI{59}{\km} long 6-mode \gls{FMF}. The output of the loop setup is amplified and forwarded to a coherent receiver array. The produced electrical signals are digitized by a 24 channel oscilloscope, followed by offline \gls{DSP}. 12 $\times$ 12 equalization is carried out using a \gls{MIMO} equalizer updated by a fully supervised \gls{LMS} algorithm. After equalization, the eigenvalues \eig{LMS} are computed at each frequency of $\mathbf{W}_{\mathrm{LMS}}$, and averaged across the signal band. The $\mathrm{SINR}_{i}$ is computed for each  of the 12 equalized data streams. The implementation penalty is computed in back-to-back as $\mathrm{SNR_{imp}}$ = 18.6 dB.
 
 After ANN training by 46,035 labelled samples, the ANN-based method is applied to experimental traces corresponding to transmission distances between 59 km and 5,900 km.
 
 \begin{figure*}[t]
\centering
    	\includegraphics[width=1\linewidth]{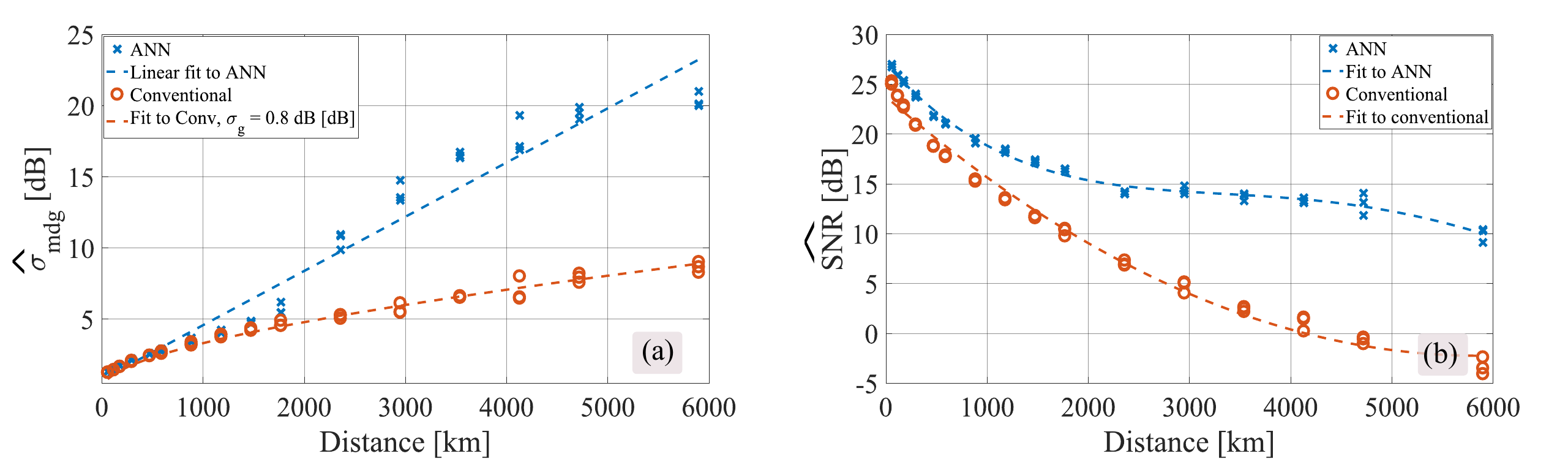}%
\caption{{Experimental long-haul 6-mode case study. (a) Estimated $\sigma_{\mathrm{mdg}}$ as a function of the transmission distance. (b) Estimated SNR as a function of the transmission distance.}}
\label{fig:nn_bell_labs}
\end{figure*}

\begin{figure*}[t]
\centering
    	\includegraphics[width=1\linewidth]{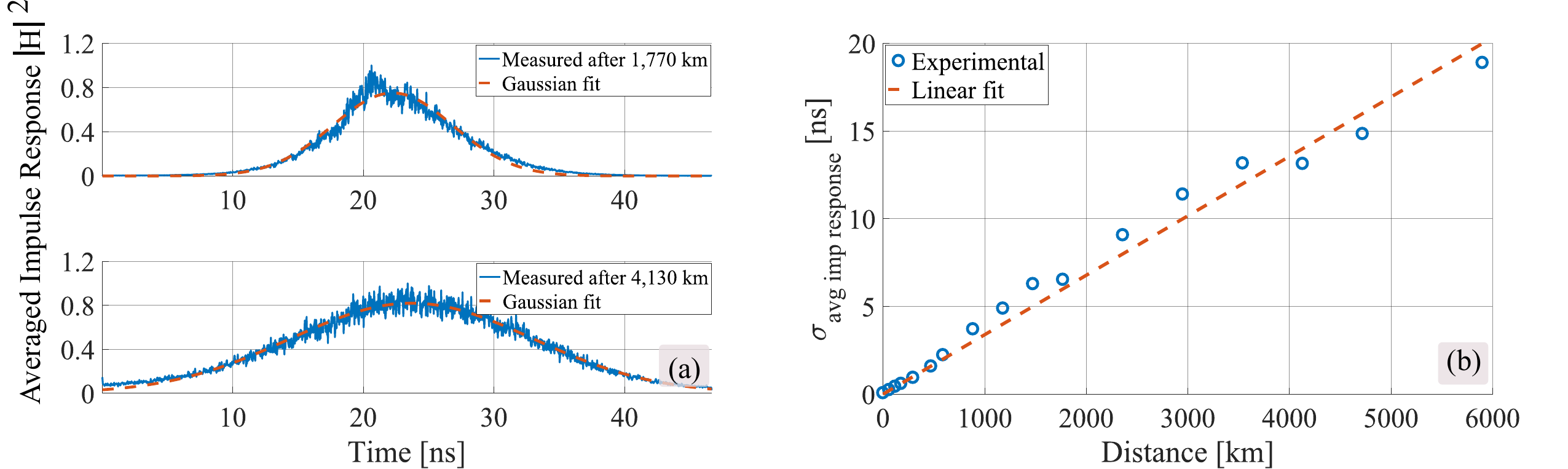}%
\caption{{(a) Averaged impulse response after 1,770 km (Top) and 4,130 km (Bottom). (b) Standard deviation of the averaged impulse response as a function of the transmission distance.}}
\label{fig:avg_impulse_response}
\end{figure*}

\subsection{Experimental long-haul 6-mode case study results}
\label{subsection:results_casestudy}

 \cref{fig:nn_bell_labs} shows $\sigma_{\mathrm{mdg}}$ and $\mathrm{SNR}$ estimated in the long-haul case study.
\cref{fig:nn_bell_labs}(a) shows the estimated $\sigma_{\mathrm{mdg}}$ as a function of the transmission distance. The orange circles correspond to  $\sigma_{\mathrm{mdg}}$ estimated by the conventional method. The dashed orange line fits the experimental data to \cref{Eq:theosigmamdg} with a per-span MDG of $\sigma_{\mathrm{g}}=\SI{0.8}{\decibel}$. The ANN-based estimates are shown by the blue crosses, indicating a large deviation with respect to the conventional method. The approximately linear increase of the  $\sigma_{\mathrm{mdg}}$ estimated by the ANN suggests a possible weakly-coupled SDM transmission \cite{lobato2012impact}. 
\cref{fig:nn_bell_labs}(b) shows the estimated SNR as a function of the transmission distance. As expected, the conventional technique results are substantially lower than those obtained by the ANN-based method. As the conventional method neglects the detrimental effects of MDG, it tends to underestimate the actual SNR. The dashed curves are polynomial fitting functions indicating the trend of the SNR with the transmission distance.

 To further investigate the weak coupling hypothesis, we also evaluate the channel delay spread.  In general, weakly-coupled transmission leads to a linear increase of the channel delay spread \cite{ho2011statistics}, \cite{antonelli2015delay}.  \cref{fig:avg_impulse_response}(a) shows the averaged impulse response computed as the average of the 144 intensity matrices obtained from the MIMO equalizer after 1,770 km and 4,130 km. The dashed red curve is a Gaussian fit whose standard deviation provides a metric for evaluating the total delay spread. \cref{fig:avg_impulse_response}(b) shows the standard deviation of the averaged impulse response as a function of the transmission distance. The approximately linear increase in the equalizer impulse response corroborates the hypothesis of weak coupling \cite{ho2011statistics}.

\section{Discussion}

In Section V, we observed considerable differences between ANN-based and conventional estimation methods for $\sigma_{\textrm{mdg}}$ and SNR in long-haul transmission. We conjectured that the very high values of MDG estimated by the ANN appeared because of a potential linear accumulation of MDG in the recirculation loop. To further understand the problem we attempt in this section to reproduce by simulation the results observed in Section V.

The simulated transmitter generates 12 16-QAM symbol sequences at \SI{30}{\giga\baud}. The sequences are processed by \gls{RRC} shaping filters and converted to the optical domain by an MZM model.
The channel model generates 12$\times$12 channel transfer matrices $\mathbf{H}$ using the analytical multisection model presented in \cite{ho2011mode}. The per-span MDG $\sigma_{g}$ is set to 1.5 dB. The transmission distance is varied from 1 to 100 59-km spans, yielding $\sigma_{\mathrm{mdg}}$ from 0.5 dB to 22 dB. Such high MDG would severely impair the transmission capacity (according to \cite{mello2019impact}, effective SNR losses higher than 1 dB are expected for $\sigma_{\mathrm{mdg}}>$ 3-4 dB). The SNR after the first span is set to 26.83 dB, and then decreased considering noise accumulation generated by amplifiers with 9-dB noise figure. 
The received sequence is fed into a coherent receiver model. The digital signals are processed by a DSP chain composed of an static equalizer and a 12$\times$12 MIMO equalizer updated by the fully-supervised \gls{LMS} algorithm.

Three $\sigma_{\mathrm{mdg}}$ estimation methods are evaluated.
The LMS-based conventional method transmits symbols over the channel matrix $\mathbf{H}$. At the receiver, the transfer matrix of the dynamic LMS MIMO equalizer is used to estimate the eigenvalues $\lambda^{2}_{i_{\mathrm{LMS}}}$ and $\sigma_{\mathrm{mdg}}$.
The LMS-based correction factor methods applies the correction factor proposed and validated in \cite{ospina2020dsp,van2020experimental,ospina2020mode} to the DSP-estimated eigenvalues before estimating  $\sigma_{\mathrm{mdg}}$.
The \gls{ANN} estimator uses the same structure and training data as in the long-haul case study in Section V.


\begin{figure*}[t]
\centering
    	\includegraphics[width=1\linewidth]{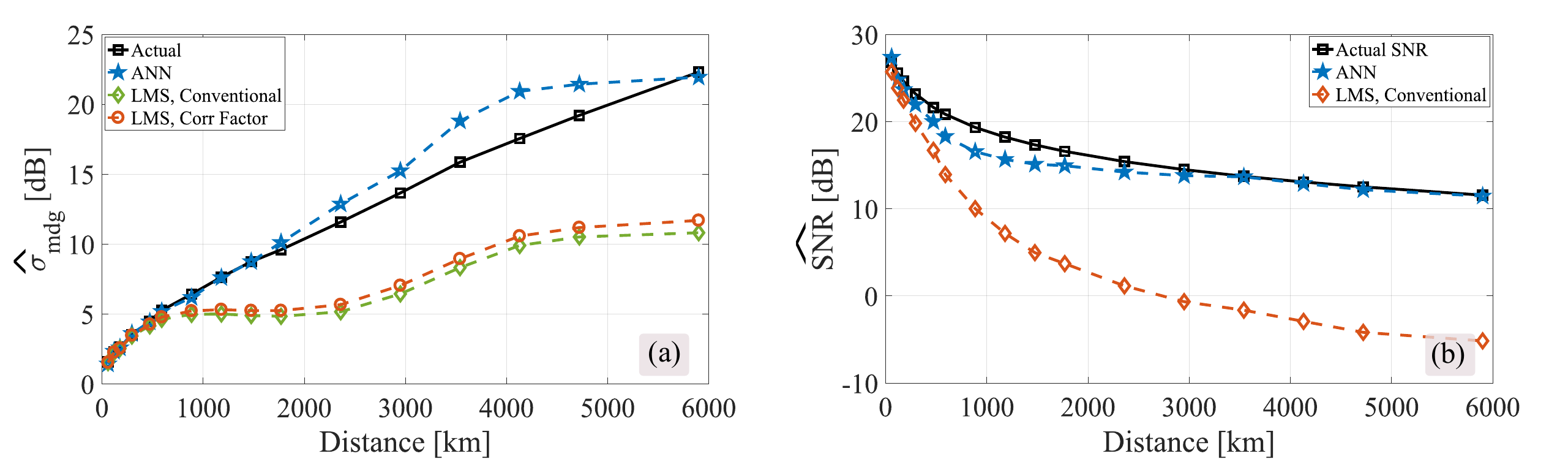}%
\caption{\small{(a) Estimated $\sigma_{\mathrm{mdg}}$ as a function of the transmission distance with conventional, correction factor, and ANN methods. (b) Estimated SNR as a function of the transmission distance with conventional and ANN methods.} }
\label{fig:semi_conv_cfactor_ann_regime}
\end{figure*}

The results are shown in \cref{fig:semi_conv_cfactor_ann_regime}(a). The solid black line corresponds to the actual $\sigma_{\mathrm{mdg}}$ estimated from $\mathbf{H}$. The LMS-based conventional method provides accurate estimates with an estimation error less than 0.5 dB up to $\sigma_{\mathrm{mdg}}$ = 4.5 dB. After this point, the method starts to significantly underestimate the MDG, reaching two plateaus.
The LMS-based technique with correction factor slightly improves the estimation quality. The correction factor provides a low correction capability at the beginning of the link because the SNR is relatively high. At the end of the link, the accumulated MDG is so high that the equalizer coefficients diverge from the MMSE coefficients. 
 The estimates provided by the ANN-based estimator accurately track the actual MDG up to $\sigma_{\mathrm{mdg}} =$  10 dB. For $\sigma_{\mathrm{mdg}} >$  10 dB, the ANN estimator slightly overestimates $\sigma_{\mathrm{mdg}}$. Although the two plateaus are not observed in the experimental data, we believe the trends are fairly reproduced.

We also estimate the SNR using the conventional (after LMS equalization) and ANN-based methods. The results are shown in  \cref{fig:semi_conv_cfactor_ann_regime}(b). As expected, the SNR is underestimated by the conventional method because of the strong MDG added to the link. The ANN-based method offers more accurate estimates, also in reasonable agreement with the experimental results in Section V.

Finally, it should be noted that the entire simulation and experimental study was carried out under the assumption of balanced and spatially white noise. This assumption should hold in real-life long-haul links, which are in fact the systems that suffer most from MDG \cite{ho2011mode}.
The effectiveness of the ANN in scenarios with noise correlation or SNR imbalances is left for a further study.

    \glsreset{SDM}
    \glsreset{MDL}
    \glsreset{MDG}
    \glsreset{SNR}
    \glsreset{MIMO}
    \glsreset{MMSE}
    \glsreset{ASE}
    \glsreset{DSP}
    \glsreset{ANN}

    \section{Conclusion}
    In \gls{SDM} systems with coupled channels, the interaction of \gls{MDG} and \gls{ASE} fundamentally constrain  the channel capacity and transmission distance.
     In these systems, accurate MDG and \gls{SNR} estimation is mandatory for an adequate link assessment and troubleshooting. Conventional estimation methods present performance limitations in certain conditions of MDG and SNR. In this paper, we investigate an \gls{ANN}-based method to estimate MDG and SNR in SDM systems with coupled channels based on features extracted after \gls{DSP}. The proposed method is validated in an experimental short-reach 3-mode transmission setup with polarization multiplexing. After validation, the ANN-based method is applied to a case study consisting of an experimental long-haul 6-mode transmission link with polarization multiplexing. The results suggest that the ANN-based method can largely exceed the performance provided by conventional methods in scenarios of high accumulated MDG, as in long-haul links with weak mode coupling.


\IEEEtriggeratref{27}


%
    \bibliographystyle{IEEEtran}
    \bibliography{refR}

\begin{thebibliography}{10}
\providecommand{\url}[1]{#1}
\csname url@samestyle\endcsname
\providecommand{\newblock}{\relax}
\providecommand{\bibinfo}[2]{#2}
\providecommand{\BIBentrySTDinterwordspacing}{\spaceskip=0pt\relax}
\providecommand{\BIBentryALTinterwordstretchfactor}{4}
\providecommand{\BIBentryALTinterwordspacing}{\spaceskip=\fontdimen2\font plus
\BIBentryALTinterwordstretchfactor\fontdimen3\font minus
  \fontdimen4\font\relax}
\providecommand{\BIBforeignlanguage}[2]{{%
\expandafter\ifx\csname l@#1\endcsname\relax
\typeout{** WARNING: IEEEtran.bst: No hyphenation pattern has been}%
\typeout{** loaded for the language `#1'. Using the pattern for}%
\typeout{** the default language instead.}%
\else
\language=\csname l@#1\endcsname
\fi
#2}}
\providecommand{\BIBdecl}{\relax}
\BIBdecl

\bibitem{ryf2019coupled}
R.~Ryf, J.~C. Alvarado-Zacarias, S.~Wittek \emph{et~al.}, ``Coupled-core
  transmission over 7-core fiber,'' in \emph{Optical Fiber Communication
  Conference}.\hskip 1em plus 0.5em minus 0.4em\relax Optical Society of
  America, 2019, pp. Th4B--3.

\bibitem{fontaine201530}
N.~K. Fontaine, R.~Ryf, H.~Chen \emph{et~al.}, ``30 {X} 30 {MIMO} transmission
  over 15 spatial modes,'' in \emph{Optical Fiber Communication
  Conference}.\hskip 1em plus 0.5em minus 0.4em\relax Optical Society of
  America, 2015, pp. Th5C--1.

\bibitem{van20201ecoc}
M.~van~den Hout, S.~van~der Heide, J.~van Weerdenburg \emph{et~al.}, ``1
  {T}bit/s/$\lambda$ transmission over a 130 km link consisting of graded-index
  50 $\mu$m core multi-mode fiber and 6{LP} few-mode fiber,'' in \emph{2020
  European Conference on Optical Communications (ECOC)}.\hskip 1em plus 0.5em
  minus 0.4em\relax IEEE, 2020, pp. 1--4.

\bibitem{van2018138}
J.~van Weerdenburg, R.~Ryf, J.~C. Alvarado-Zacarias \emph{et~al.}, ``138-{T}b/s
  mode-and wavelength-multiplexed transmission over six-mode graded-index
  fiber,'' \emph{Journal of Lightwave Technology}, vol.~36, no.~6, pp.
  1369--1374, 2018.

\bibitem{rademacher2018long}
G.~Rademacher, R.~Ryf, N.~K. Fontaine \emph{et~al.}, ``Long-haul transmission
  over few-mode fibers with space-division multiplexing,'' \emph{Journal of
  Lightwave Technology}, vol.~36, no.~6, pp. 1382--1388, 2018.

\bibitem{soma201710}
D.~Soma, Y.~Wakayama, S.~Beppu \emph{et~al.}, ``10.16 peta-bit/s dense
  {SDM/WDM} transmission over low-{DMD} 6-mode 19-core fibre across {C}+{L}
  band,'' in \emph{2017 European Conference on Optical Communication
  (ECOC)}.\hskip 1em plus 0.5em minus 0.4em\relax IEEE, 2017, pp. 1--3.

\bibitem{ho2011mode}
K.-P. Ho and J.~M. Kahn, ``Mode-dependent loss and gain: statistics and effect
  on mode-division multiplexing,'' \emph{Optics express}, vol.~19, no.~17, pp.
  16\,612--16\,635, 2011.

\bibitem{winzer2011mimo}
P.~J. Winzer and G.~J. Foschini, ``{MIMO} capacities and outage probabilities
  in spatially multiplexed optical transport systems,'' \emph{Optics express},
  vol.~19, no.~17, pp. 16\,680--16\,696, 2011.

\bibitem{8918466}
D.~A.~A. {Mello}, H.~{Srinivas}, K.~{Choutagunta} \emph{et~al.}, ``Impact of
  polarization- and mode-dependent gain on the capacity of ultra-long-haul
  systems,'' \emph{Journal of Lightwave Technology}, vol.~38, no.~2, pp.
  303--318, 2020.

\bibitem{van2017138}
J.~van Weerdenburg, R.~Ryf, J.~C. Alvarado-Zacarias \emph{et~al.}, ``138
  {T}bit/s transmission over 650 km graded-index 6-mode fiber,'' in \emph{2017
  European Conference on Optical Communication (ECOC)}.\hskip 1em plus 0.5em
  minus 0.4em\relax IEEE, 2017, pp. 1--3.

\bibitem{rademacher202010}
G.~Rademacher, B.~J. Puttnam, R.~S. Luís \emph{et~al.}, ``10.66 {P}eta-{B}it/s
  transmission over a 38-core-three-mode fiber,'' in \emph{Optical Fiber
  Communication Conference}.\hskip 1em plus 0.5em minus 0.4em\relax Optical
  Society of America, 2020, pp. Th3H--1.

\bibitem{ospina2020dsp}
R.~S. Ospina, C.~Okonkwo, and D.~A. Mello, ``{DSP}-based mode-dependent loss
  and gain estimation in coupled {SDM} transmission,'' in \emph{Optical Fiber
  Communication Conference}.\hskip 1em plus 0.5em minus 0.4em\relax Optical
  Society of America, 2020, pp. W2A--47.

\bibitem{faruk2017digital}
M.~S. Faruk and S.~J. Savory, ``Digital signal processing for coherent
  transceivers employing multilevel formats,'' \emph{Journal of Lightwave
  Technology}, vol.~35, no.~5, pp. 1125--1141, 2017.

\bibitem{ospina2020mode}
R.~S. Ospina, M.~Van~den Hout, J.~C. Alvarado-Zacarias \emph{et~al.},
  ``Mode-dependent loss and gain estimation in {SDM} transmission based on
  {MMSE} equalizers,'' \emph{Journal of Lightwave Technology}, vol.~39, no.~7,
  pp. 1968--1975, 2020.

\bibitem{van2020experimental}
M.~van~den Hout, R.~S. Ospina, S.~van~der Heide \emph{et~al.}, ``Experimental
  validation of {MDL} emulation and estimation techniques for {SDM}
  transmission systems,'' in \emph{2020 European Conference on Optical
  Communications (ECOC)}.\hskip 1em plus 0.5em minus 0.4em\relax IEEE, 2020,
  pp. 1--4.

\bibitem{essiambre2010capacity}
R.-J. Essiambre, G.~Kramer, P.~J. Winzer \emph{et~al.}, ``Capacity limits of
  optical fiber networks,'' \emph{Journal of Lightwave Technology}, vol.~28,
  no.~4, pp. 662--701, 2010.

\bibitem{saif2020machine}
W.~Saif, M.~A. Esmail, A.~Ragheb \emph{et~al.}, ``Machine learning techniques
  for optical performance monitoring and modulation format identification: A
  survey,'' \emph{IEEE Communications Surveys \& Tutorials}, 2020.

\bibitem{saif2020optical}
W.~S. Saif, A.~M. Ragheb, T.~A. Alshawi \emph{et~al.}, ``Optical performance
  monitoring in mode division multiplexed optical networks,'' \emph{Journal of
  Lightwave Technology}, vol.~39, no.~2, pp. 491--504, 2020.

\bibitem{ospina2021ann}
R.~S.~B. Ospina, M.~van~den Hout, S.~van~der Heide \emph{et~al.},
  ``{Neural-network-based MDG and Optical SNR Estimation in SDM
  Transmission},'' 2021, arXiv preprint
  \href{https://arxiv.org/abs/2104.06803}{arXiv:2104.06803}, 2021, accepted for
  presentation at OFC 2021.

\bibitem{kim2008performance}
N.~Kim, Y.~Lee, and H.~Park, ``Performance analysis of {MIMO} system with
  linear {MMSE} receiver,'' \emph{IEEE Transactions on Wireless
  Communications}, vol.~7, no.~11, pp. 4474--4478, 2008.

\bibitem{mckay2009achievable}
M.~R. McKay, I.~B. Collings, and A.~M. Tulino, ``Achievable sum rate of {MIMO
  MMSE} receivers: A general analytic framework,'' \emph{IEEE Transactions on
  Information Theory}, vol.~56, no.~1, pp. 396--410, 2009.

\bibitem{hayashi2017record}
T.~Hayashi, Y.~Tamura, T.~Hasegawa \emph{et~al.}, ``Record-low spatial mode
  dispersion and ultra-low loss coupled multi-core fiber for ultra-long-haul
  transmission,'' \emph{Journal of Lightwave Technology}, vol.~35, no.~3, pp.
  450--457, 2017.

\bibitem{wautelet2007comparison}
X.~Wautelet, C.~Herzet, A.~Dejonghe \emph{et~al.}, ``Comparison of em-based
  algorithms for {MIMO} channel estimation,'' \emph{IEEE Transactions on
  communications}, vol.~55, no.~1, pp. 216--226, 2007.

\bibitem{das2012snr}
A.~Das and B.~D. Rao, ``{SNR} and noise variance estimation for {MIMO}
  systems,'' \emph{IEEE Transactions on Signal processing}, vol.~60, no.~8, pp.
  3929--3941, 2012.

\bibitem{kingma2014adam}
D.~P. Kingma and J.~Ba, ``Adam: A method for stochastic optimization,''
  \emph{arXiv preprint arXiv:1412.6980}, 2014.

\bibitem{sillard201650}
P.~Sillard, D.~Molin, M.~Bigot-Astruc \emph{et~al.}, ``50 {$\mu$}m multimode
  fibers for mode division multiplexing,'' \emph{Journal of Lightwave
  Technology}, vol.~34, no.~8, pp. 1672--1677, 2016.

\bibitem{velazquez2018scaling}
A.~M. Velázquez~Benítez, J.~E. Antonio~López, J.~C. Alvarado~Zacarías
  \emph{et~al.}, ``Scaling photonic lanterns for space-division multiplexing,''
  \emph{Scientific reports}, vol.~8, no.~1, pp. 1--9, 2018.

\bibitem{van2014ultra}
R.~Van~Uden, R.~A. Correa, E.~A. Lopez \emph{et~al.}, ``Ultra-high-density
  spatial division multiplexing with a few-mode multicore fibre,'' \emph{Nature
  Photonics}, vol.~8, no.~11, p. 865, 2014.

\bibitem{da2016widely}
E.~P. da~Silva and D.~Zibar, ``{Widely linear equalization for IQ imbalance and
  skew compensation in optical coherent receivers},'' \emph{Journal of
  Lightwave Technology}, vol.~34, no.~15, pp. 3577--3586, 2016.

\bibitem{lobato2012impact}
A.~Lobato, F.~Ferreira, M.~Kuschnerov \emph{et~al.}, ``Impact of mode coupling
  on the mode-dependent loss tolerance in few-mode fiber transmission,''
  \emph{Optics express}, vol.~20, no.~28, pp. 29\,776--29\,783, 2012.

\bibitem{ho2011statistics}
K.-P. Ho and J.~M. Kahn, ``Statistics of group delays in multimode fiber with
  strong mode coupling,'' \emph{Journal of lightwave technology}, vol.~29,
  no.~21, pp. 3119--3128, 2011.

\bibitem{antonelli2015delay}
C.~Antonelli, A.~Mecozzi, and M.~Shtaif, ``The delay spread in fibers for {SDM}
  transmission: dependence on fiber parameters and perturbations,''
  \emph{Optics express}, vol.~23, no.~3, pp. 2196--2202, 2015.

\bibitem{mello2019impact}
D.~A.~A. {Mello}, H.~{Srinivas}, K.~{Choutagunta} \emph{et~al.}, ``Impact of
  polarization- and mode-dependent gain on the capacity of ultra-long-haul
  systems,'' \emph{Journal of Lightwave Technology}, vol.~38, no.~2, pp.
  303--318, 2020.

\end{thebibliography}
%









\end{document}